\documentclass[]{mn2e}

\usepackage{psfig,subfigure,mncite}

\begin{document}
 

\title[Variable differential rotation on AB Dor]
{Doin' the twist: Secular changes in the surface 
differential rotation on AB Doradus}

\author
[A. Collier~Cameron, J.-F. Donati]
{A. Collier Cameron$^1$
\thanks{E-mail: andrew.cameron@st-and.ac.uk},
J.-F. Donati$^2$
\\
$^1$School of Physics and Astronomy, Univ.\ of St~Andrews, 
St~Andrews, 
Scotland KY16 9SS \\
$^2$Laboratoire d'Astrophysique, Observatiore Midi-Pyr\'{e}n\'{e}es,
Avenue E. Belin, F-31400 Toulouse, France
\\
} 

\date{Accepted 2001 November 12. Received 2001 November 12; in original
form 2001 November 6}

\maketitle

\begin{abstract} 
We present measurements of the rotation rates of individual starspots
on the rapidly rotating young K0 dwarf AB Doradus, at six epochs
between 1988 December and 1996 December.  The equatorial
rotation period of the star decreased from 0.5137 to 0.5129
days between 1988 December and 1992 January.  It then increased
steadily, attaining a value of 0.5133 days by 1996 December.  The
latitude dependence of the rotation rate mirrored the changes in the
equatorial rotation rate.  The beat period between the equatorial and
polar rotation periods dropped from 140 days to 70 days initially,
then rose steadily.  The most rigid rotation, in 1988 December,
occurred when the starspot coverage was at a maximum.  The
time-dependent part of the differential rotation is found to have
$\Delta\Omega/\Omega\simeq 0.004$, which should alter the oblateness
of the star enough to explain the period changes observed in several
close binaries via the mechanism of \scite{applegate92}.
\end{abstract}


\section{Introduction}

The back-reaction of Lorentz forces on the fluid circulation in the
convective zone of a rotating late-type star is of fundamental
importance in theories of stellar magnetic-field generation.  Over the
last decade or so, a consensus has emerged among dynamo theorists that
this back-reaction could have a significant influence on the
differential rotation pattern in the convective zones of the Sun and
active late-type stars (see, e.g.
\pcite{brandenburg91,moss95,kitchatinov99SoPh}).

The quasi-cyclic orbital period changes observed among short-period
binaries with magnetically-active components provide further indirect
evidence that substantial changes in differential rotation could be
taking place on active binary components.  Work by
\scite{applegate92}, \scite{lanza98} and \scite{lanza99} suggests that
small, magnetically-modulated changes in stellar differential rotation
will alter the gravitational quadrupole moment of the active star. 
This can be sufficient to produce orbital period changes as large as the
fractional $\delta P/P\simeq 1.5\times 10^{-4}$ observed in HR 1099 by
\scite{donati99hr1099}.  

In this paper we report observations of long-term changes in surface
differential rotation on the young, rapidly rotating K0 dwarf AB
Doradus, providing independent support for this theory.  We apply a
new method for determining the latitudes and rotation rates of
individual starspots \cite{cameron2001diffrot} to six sets of archival
time-resolved echelle spectroscopy of AB Dor, spanning the period from
1988 December to 1996 December.

\section{Observations and data reduction}

The details of the instruments and observing procedures used to secure
the six data sets have been published elsewhere, so we give only a
brief summary in Table~\ref{tab:obser}.

\begin{table*}
\caption[]{Summary of observing runs. Details of the observations are
given in the papers listed. The phase ranges in the fourth column are
those in which the complete signatures of individual spots could be 
observed  on at least two nights of the run observing concerned.} 

\begin{tabular}{lcrcl}
UT Dates & Telescope/Instrument & No of & Phase coverage & References\\
         &                      & lines &                & \\
&&&\\
1988 Dec 16, 19		& 3.9-m AAT/UCLES & 139 & 0.09:0.40& \scite{cameron90masses}\\
1988 Dec 21, 23 	& 3.6-m ESO/CASPEC & 326 & 0.0:0.38, 0.94:1.00&\scite{cameron90masses}\\
1992 Jan 18, 19, 20	& 3.9-m AAT/UCLES & 1134 & 0.08:0.62 & \scite{cameron94doppler}\\
1993 Nov 23, 24, 25	& 3.9-m AAT/UCLES & 1566 & 0.40:0.85 & \scite{unruh95doppler}\\
1994 Nov 15, 16, 17	& 4.0-m CTIO/echelle& 619 & 0.04:0.18, 0.28:0.63& \scite{cameron99musicos}\\
1995 Dec 07, 11		& 3.9-m AAT/UCLES & 1936 & 0.55:0.95 & \scite{donati97doppler}\\
1996 Dec 23, 25, 27, 29	& 3.9-m AAT/UCLES & 1964 & 0.50:1.00 & \scite{donati99doppler}\\
\end{tabular}
\label{tab:obser}
\end{table*}

The observations from the three earliest runs were re-extracted from
the raw data using the Starlink ECHOMOP optimal extraction routines,
to ensure consistency with the later data.  The spectra from all 6
years included the H$\alpha$ region, in which numerous weak, narrow
telluric lines of H$_2$O and O$_2$ are present.  We used spectral
subtraction \cite{cameron2001upsand} to isolate the travelling
distortions produced by starspots in the mean profile, and
least-squares deconvolution \cite{donati97zdi} to stack up the
residual profile information in the large number of known photospheric
lines recorded in the echellograms.  The number $N$ of lines used in
each year ranges from 140 to 2000, and is given in
Table~\ref{tab:obser}, column 3.  Since the spectra in all years were
exposed to a signal-to-noise (S:N) ratio of 100 to 120, the S:N of the
deconvolved profiles scales approximately with $\sqrt{N}$.

The deconvolved profiles were placed in the heliocentric reference
frame, to an accuracy better than 100 m s$^{-1}$, using the telluric
lines as velocity references during the deconvolution procedure.  The
resulting time-series of residual profiles were subjected to a
matched-filter analysis, yielding measures of the spot area, radial
velocity amplitude about the stellar centre of mass, and rotation
period, together with estimates of their uncertainties.  More detailed
descriptions of the deconvolution, spectral subtraction and
matched-filter analysis procedures are given by
\scite{cameron2001upsand} and \scite{cameron2001diffrot}.  As in this
earlier paper, we used a limb-darkening coefficient $u=0.77$ and
$v\sin i = 91$ km s$^{-1}$ in constructing the matched filter.  

\section{Results}

The rotation periods of the spots detected in each season's data are
plotted against their rotational velocity amplitudes $K =
\Omega(\theta)R_{\star}\cos\theta\sin i$ in the left-hand panels of
Figure~\ref{fig:diffrot}.  Here $\Omega(\theta)$ is the rotation rate
at latitude $\theta$, and $R_{\star}$ is the stellar radius.

In the 1995 and 1996 seasons, outlying points caused by aliasing
between closely-spaced pairs of spots observed several days apart can
be seen \cite{cameron2001diffrot}.  These were excluded from the next
stage of the analysis.

The stellar radial velocity $v_{r}\simeq 32.5$ km~s$^{-1}$ and $v\sin
i=91$ km~s$^{-1}$are both uncertain by $\pm 1$ km~$s^{-1}$ or so.  The
systematic errors in the derived quantities $K$ and $P$ due to these
uncertainties are smaller than the $1\sigma$ error bars shown in
Figure~\ref{fig:diffrot}.  

As \scite{cameron2001diffrot} noted, there is an intrinsic scatter of the
individual spot rotation rates about the mean differential rotation
pattern. The rms magnitude of this scatter is equivalent to an additional
uncertainty of $\pm 0.00017$ day in the rotation period of each spot.
After adding this error in quadrature to the measured uncertainty in each
spot's period, we fitted a differential rotation law of the form 
\begin{equation}
\Omega(\theta) = \Omega_{\mbox{equator}} - \Omega_{\mbox{beat}}\sin^{2}\theta
\label{eq:omega}
\end{equation}
to each season's data.  The optimal solutions for the equatorial
rotation period $P_{\mbox{equator}}=2\pi/\Omega_{\mbox{equator}}$ and
the equator-pole beat period
$P_{\mbox{beat}}=2\pi/\Omega_{\mbox{beat}}$ are shown together with
their 68.3\% and 95.4\% confidence regions in the right-hand panels of
Fig.~\ref{fig:diffrot}.  The inner contour is the locus
$\chi^{2}=\chi^{2}_{\mbox{min}}+1.0$, whose extremities in
$P_{\mbox{equator}}$ and $P_{\mbox{beat}}$ give the one-dimensional
$1\sigma$ error bars on the two periods.  The outer contours, at
$\chi^{2}=\chi^{2}_{\mbox{min}}+2.3$ and 6.17, contain 68.3\%\ and
95.4\%\ of the joint probability respectively. 
	
\begin{figure*}
	\def\subfigtopskip{4pt}
	\def\subfigbottomskip{4pt}
	\def\subfigcapskip{2pt}
	\centering
	\begin{tabular}{ll}
    	\subfigure[]{
			\label{fig:diffrot88} 
			\psfig{figure=fig01a.eps,bbllx=70pt,bblly=65pt,bburx=512pt,bbury=380pt,height=6.9cm}
			} &
		\subfigure[]{
			\label{fig:contours88} 
			\psfig{file=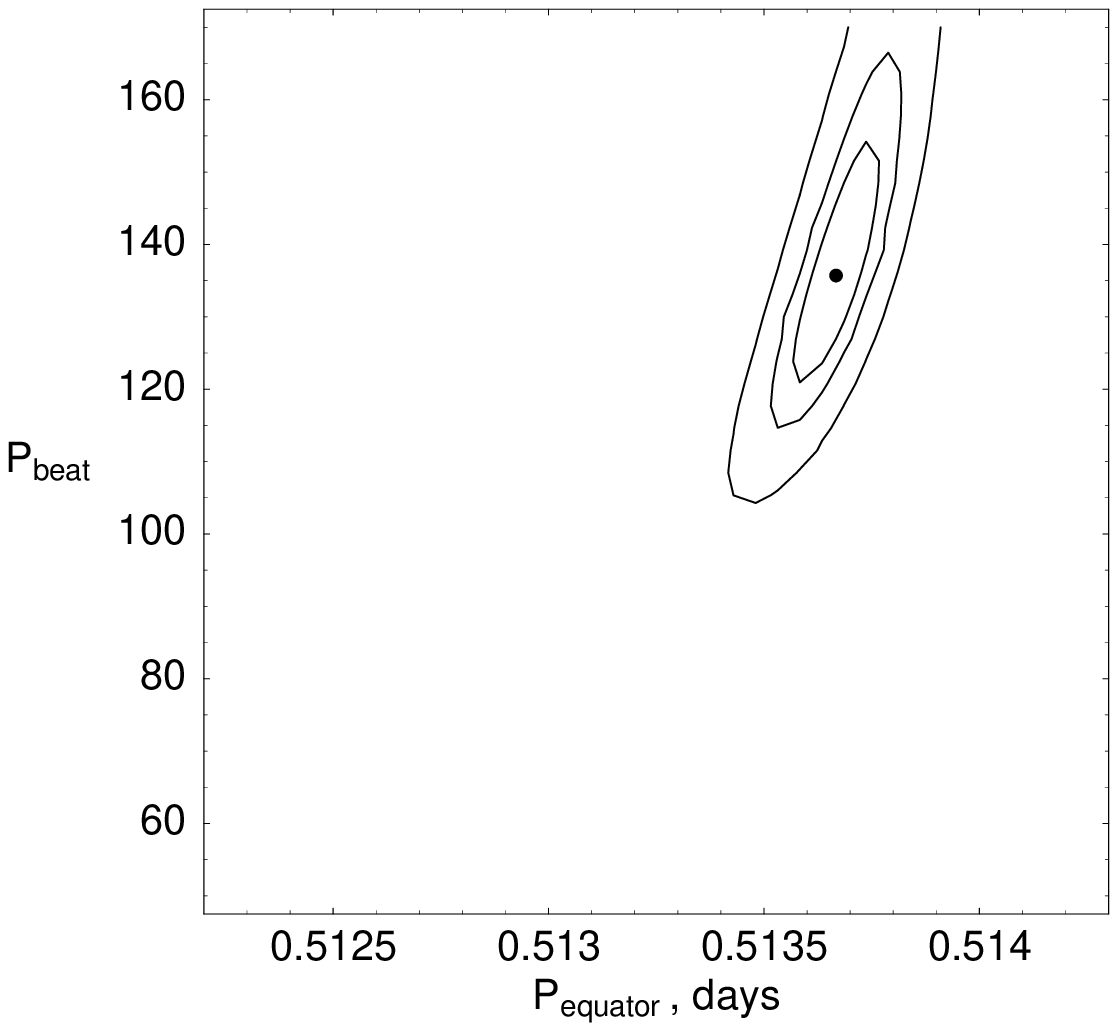,height=6.9cm}
			} \\
    	\subfigure[]{
			\label{fig:diffrot92} 
			\psfig{figure=fig01c.eps,bbllx=70pt,bblly=65pt,bburx=512pt,bbury=380pt,height=6.9cm}
			} &
		\subfigure[]{
			\label{fig:contours92} 
			\psfig{file=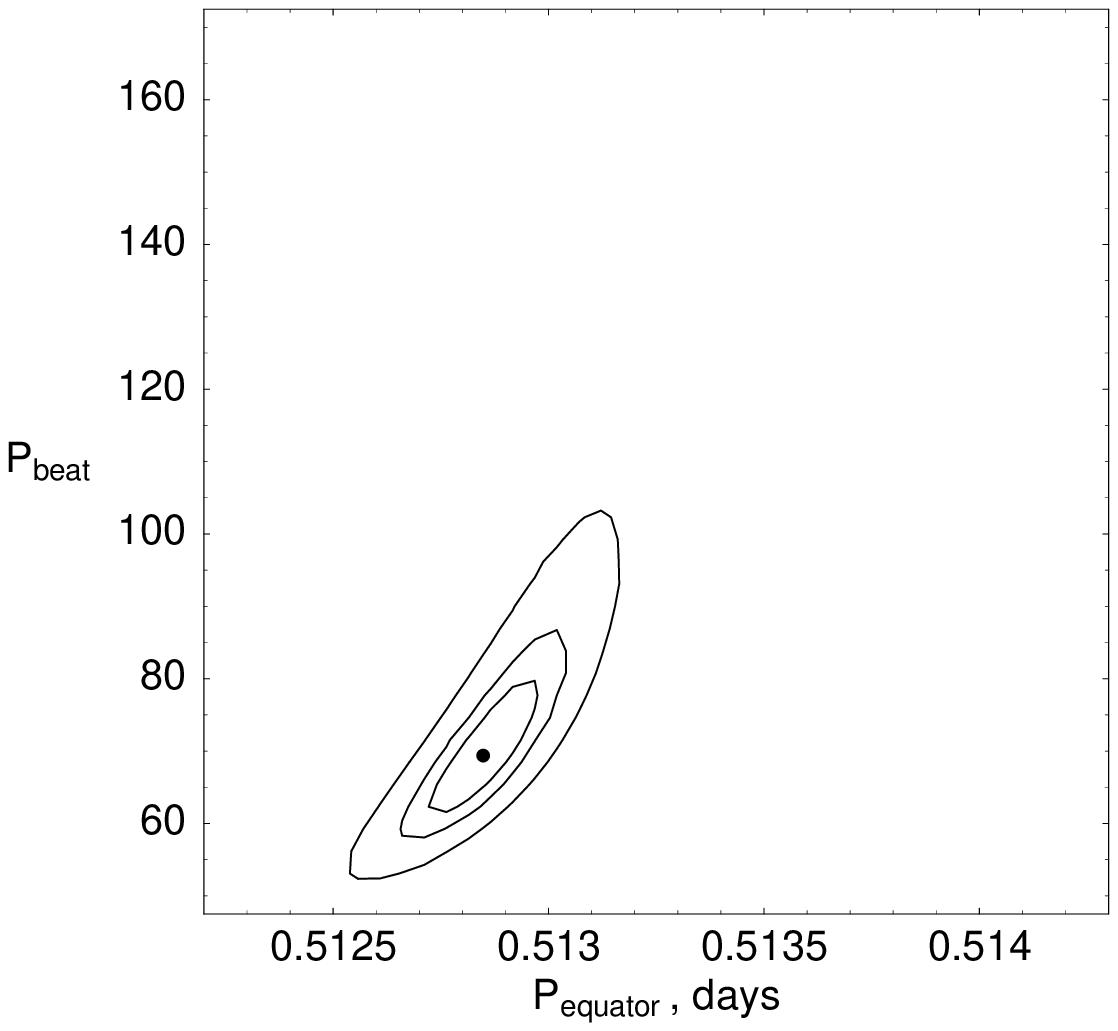,height=6.9cm}
			} \\
    	\subfigure[]{
			\label{fig:diffrot93}
			\psfig{figure=fig01e.eps,bbllx=70pt,bblly=65pt,bburx=512pt,bbury=380pt,height=6.9cm}
			} &
		\subfigure[]{
			\label{fig:contours93} 
			\psfig{file=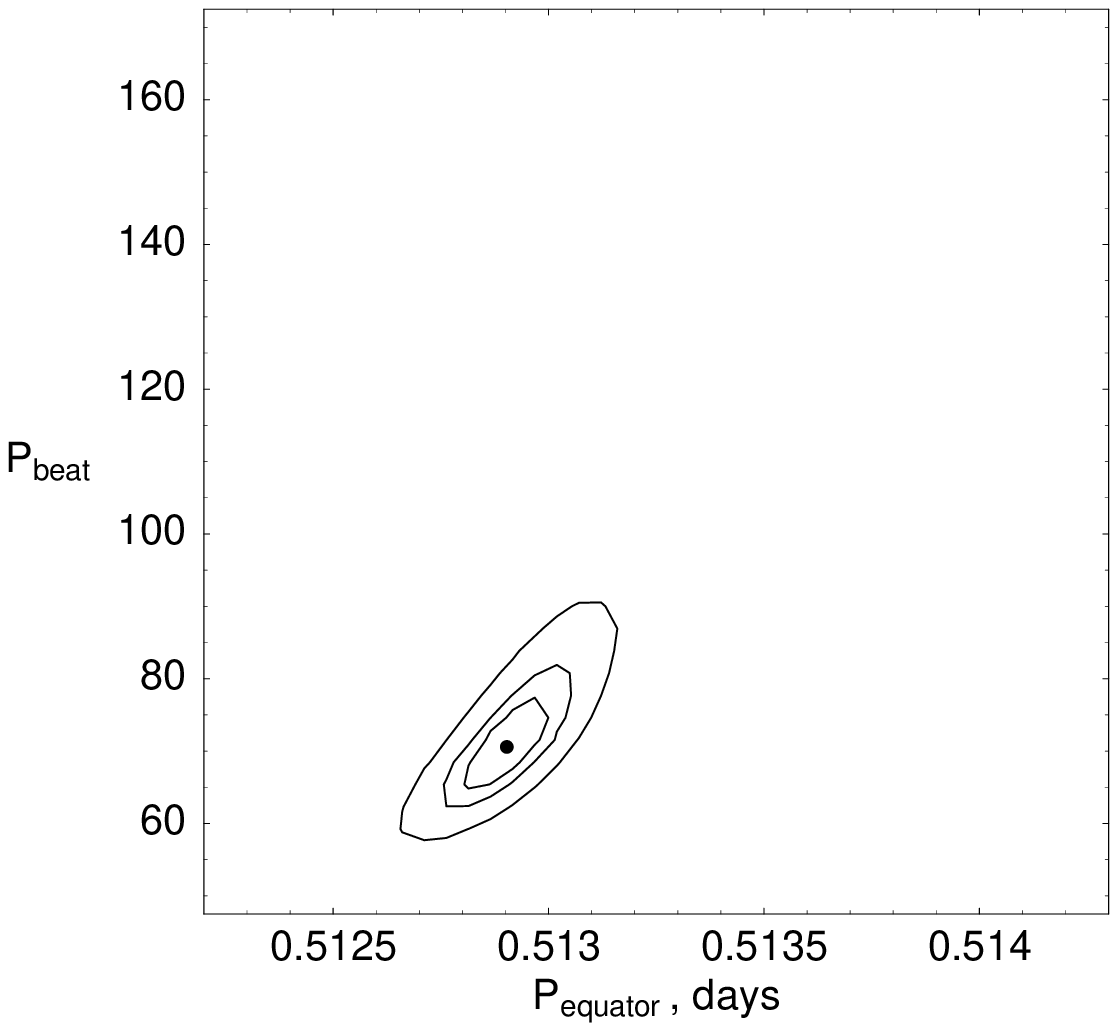,height=6.9cm}
			} \\
	\end{tabular} 
	\caption[]{The left-hand panels show rotation period $P$
	versus rotational velocity amplitude $K$ for candidate
	starspots in each year of observation.  ``Threshold'' denotes the
	$\chi^{2}$ threshold below which spurious spots have been
	rejected.  ``Phi'' denotes the phase range searched, as given
	in column 4 of Table~\ref{tab:obser}.  The fitted differential
	rotation curve is shown for each year.  The right-hand panels
	show the contours $\Delta\chi^{2}=1.0$, 2.3 and 6.17 as a
	function of equatorial rotation period and equator-pole lap
	time, both in days.}
    \label{fig:diffrot}
\end{figure*}

\begin{figure*}
	\def\subfigtopskip{4pt}
	\def\subfigbottomskip{4pt}
	\def\subfigcapskip{2pt}
	\centering
	\addtocounter{subfigure}{6}
	\begin{tabular}{ll}
    	\subfigure[]{
			\label{fig:diffrot94} 						
			\psfig{figure=fig01g.eps,bbllx=70pt,bblly=65pt,bburx=512pt,bbury=380pt,height=6.9cm}
			} &
		\subfigure[]{
			\label{fig:contours94} 
			\psfig{file=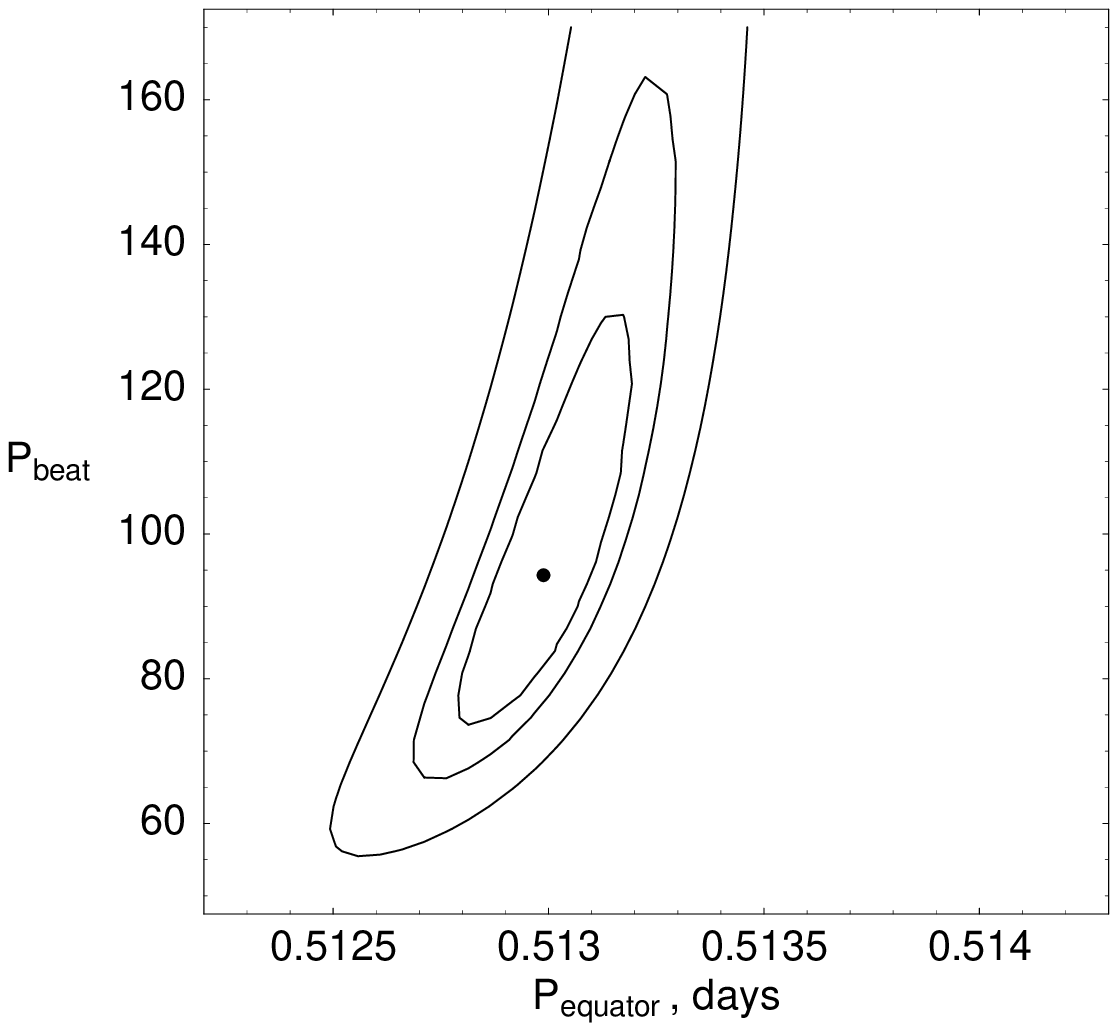,height=6.9cm}
			} \\
    	\subfigure[]{
			\label{fig:diffrot95} 						
			\psfig{figure=fig01i.eps,bbllx=70pt,bblly=65pt,bburx=512pt,bbury=380pt,height=6.9cm}
			} &
		\subfigure[]{
			\label{fig:contours95} 
			\psfig{file=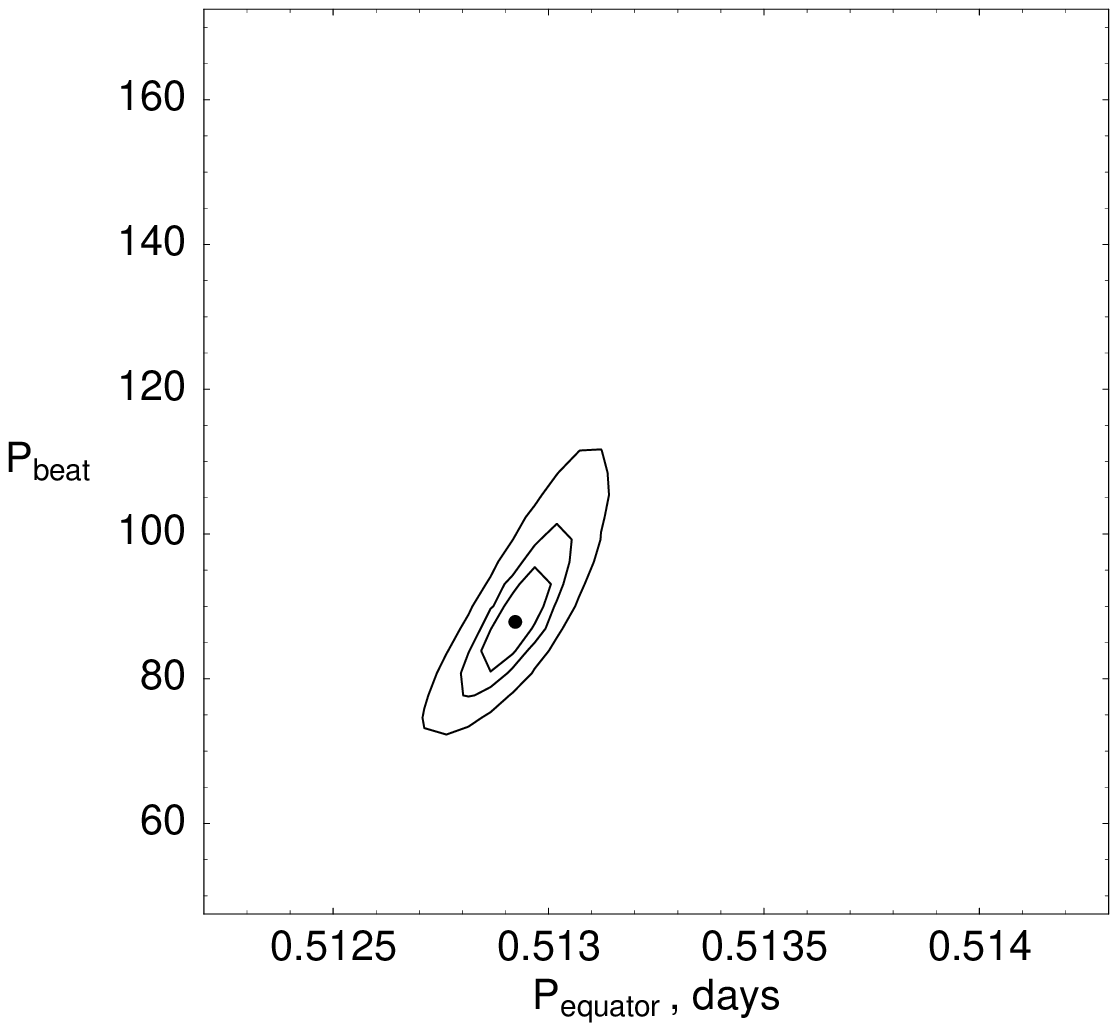,height=6.9cm}
			} \\
    	\subfigure[]{
			\label{fig:diffrot96} 			
			\psfig{figure=fig01k.eps,bbllx=70pt,bblly=65pt,bburx=512pt,bbury=380pt,height=6.9cm}
			} &
		\subfigure[]{
			\label{fig:contours96} 
			\psfig{file=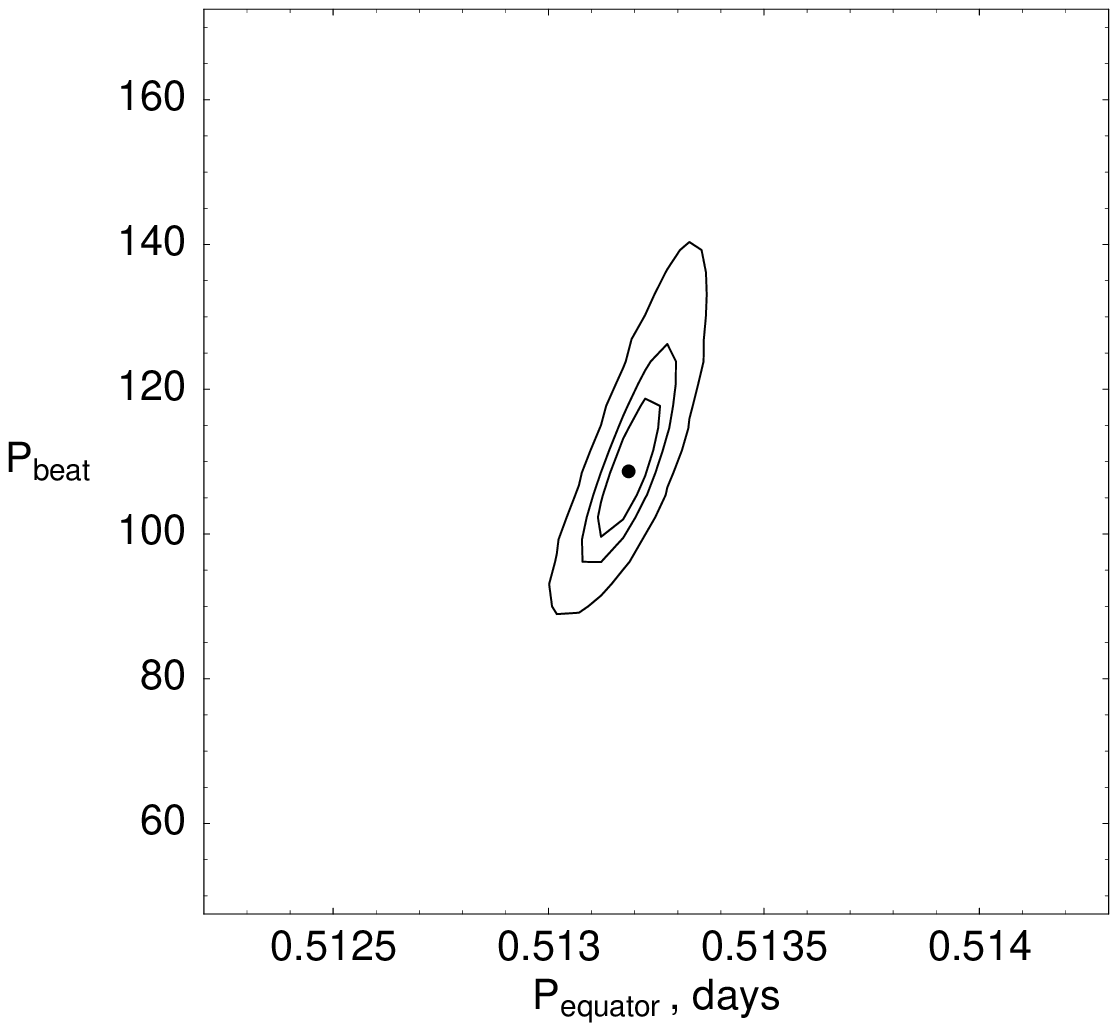,height=6.9cm}
			} \\
	\end{tabular} 
	\contcaption{}
\end{figure*}

\begin{figure}
	\def\subfigtopskip{4pt}
	\def\subfigbottomskip{4pt}
	\def\subfigcapskip{2pt}
	\centering
	\setcounter{subfigure}{0}
	\begin{tabular}{l}
    	   \subfigure[]{
		\label{fig:pbeat_jd} 						
		\psfig{figure=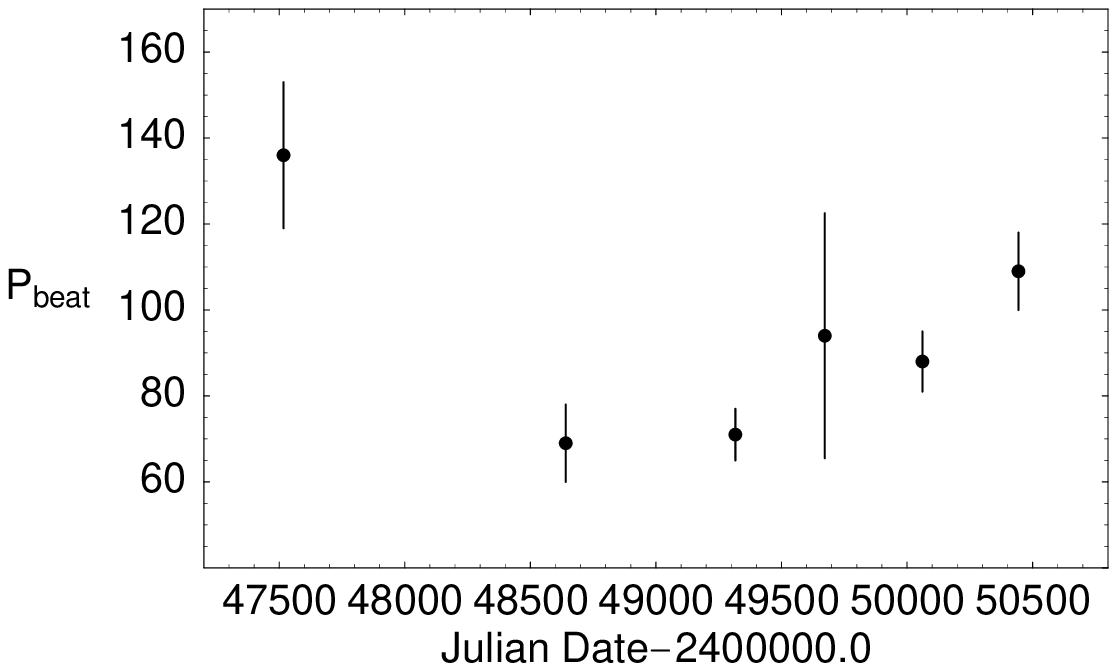,width=8.4cm}
		} \\
	   \subfigure[]{
		\label{fig:peq_jd} 
		\psfig{figure=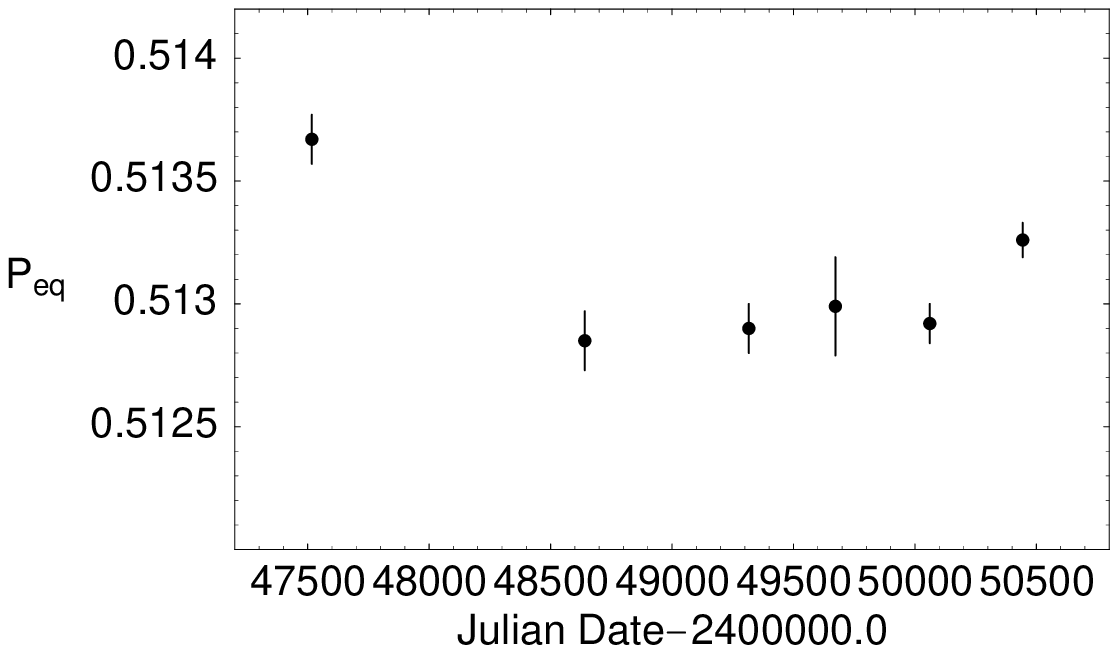,width=8.4cm}
		} \\
	\end{tabular} 	
	\caption[]{The equator-pole beat period $P_{\mbox{beat}}$
	(upper panel) and equatorial rotation period $P_{\mbox{eq}}$
	(lower panel) versus Julian date.  The data points are shown
	for (left to right) 1988 December, 1992 January, 1993
	November, 1994 November, 1995 December and 1996 December.}
    \label{fig:torsion}
\end{figure}

The fits to the six years' data are summarized in
Table~\ref{tab:fitparams} and Fig.~\ref{fig:torsion}.  The
equator-pole beat period of AB Dor displays significant variability
from one observing season to the next.  Between 1988 December and 1992
January the differential rotation rate in the star's surface layers
doubled, with the beat period falling from 140 to 70 days.  The
equatorial rate increased and the high-latitude rotation rate dropped. 
This increased shear pattern persisted into 1993.  In 1994 and 1995
the rotation rate at latitude 40$^{\circ}$ and northward increased
while the equator continued to rotate rapidly.  In 1996 the high
latitudes continued spinning up as the equator began to show signs of
slowing down.  The smallest variation in rotation period is seen at
$K\simeq 70$ km~s$^{-1}$ (latitude $\sim 40^{\circ}$), but nowhere on
the star is the rotation rate constant.

\begin{table}
	\caption[]{Differential rotation parameters derived from  
	2-parameter fits to the rotation rates of individual spots 
	in 1988 December, 1992 January, 1993 November, 1994 November,
	1995 December and 1996 December. The second column gives the number 
	of spots contributing to each fit.}
	\begin{tabular}{lrcrr}		
		JD & No of &$P_{\mbox{equator}}$ & $P_{\mbox{beat}}$ & $\chi^{2}$  \\
		   & spots   &(days) & (days)  \\
		      &                   &       &      \\
47517.1 & 12 &$0.51367\pm 0.00010$ & $ 136^{+18}_{-16}$ & 32.4 \\
48640.9 & 12 &$0.51285\pm 0.00012$ & $ 69^{+10}_{-8}$   & 11.2 \\
49316.3 & 15 &$0.51290\pm 0.00010$ & $ 71^{+6}_{-6}$    & 47.1 \\
49672.5 & 13 &$0.51299\pm 0.00020$ & $ 94^{+36}_{-21}$  & 55.2 \\
50061.2 & 16 &$0.51292\pm 0.00008$ & $ 88^{+7}_{-7}$    & 15.9 \\
50443.7 & 19 &$0.51326\pm 0.00007$ & $109^{+9}_{-9}$    & 53.3 \\
	\end{tabular}
	\label{tab:fitparams}
\end{table}

\section{Discussion and Conclusions}

We have found that the differential rotation rate of the young K0
dwarf AB Dor varies by a factor 2 on a timescale of a few years. 
While the 8-year timespan of the observations presented here is too
short to determine whether we are seeing a cyclic phenomenon, the
apparent smoothness of the increase and decline in the differential
rotation suggests that the changes occur on a timescale of order a
decade or two.  The variation resembles a wave of excess rotation,
originating at the equator in 1992 and propagating poleward.


\scite{applegate92} and \scite{lanza98} have shown that changes in
magnetic flux threading the convective zone during a stellar magnetic
cycle may alter the viscous transport of angular momentum sufficiently
to produce substantial changes in differential rotation.  AB Dor was 
rotating most rigidly in late 1988, at a time when the star's
maximum light level was depressed by 0.2 mag or so relative to
photometry secured in the early 1980s and mid-1990s \cite{kuerster97}. 
The increase in differential rotation between 1988 and early 1992
coincides with a 0.1-mag rise in the maximum brightness of the star. 
The maximum brightness of the star remained steady between 1992 and
late 1996, as the differential rotation began gradually to decrease
again.  The timescales of the long-term changes in starspot coverage
and differential rotation thus appear to be comparable, though we
cannot yet determine how closely the spot coverage and differential
rotation rate are linked.

These departures from the time-averaged differential rotation pattern
are between one and two orders of magnitude greater than the waves of
excess and deficit rotational velocity -- sometimes described as
``torsional oscillations'' -- that migrate equatorward during the
solar spot cycle \cite{howard80,labonte82}.  These departures from the
mean solar rotational velocity have amplitudes of order 5 m~s$^{-1}$. 
At the equator of AB Dor, however, a change in period from 0.5137 to
0.5128 days corresponds to a 200 m~s$^{-1}$ change in $v\sin i$.  If
we consider the full change $\Delta\Omega=0.048$ radian d$^{-1}$ in
the equator-pole beat frequency, we find that
$\Delta\Omega/\Omega\simeq 0.004$.

Such a large modulation of the surface differential rotation will
alter the star's oblateness sufficiently that, if AB Dor were in a
close binary system, it would be expected to produce observable
long-term orbital period changes.  By way of comparison, the K2V
primary of the binary system V471 Tau has a spectral type and rotation
period very similar to those of AB Dor.  \scite{applegate92} points
out that the observed 20-year, $\Delta P/P\simeq 10^{-6}$ modulation
of the orbital period in V471 Tau \cite{skillman88,ibanoglu94} would
require a variable differential rotation in the K star, with
$\Delta\Omega/\Omega = 0.0032$.  The observed changes in AB Dor's
differential rotation are of precisely the magnitude required by the
\scite{applegate92} model to explain the orbital period variations in
V471 Tau.

\section*{ACKNOWLEDGMENTS}

This paper is based on observations
made using the 3.9-m Anglo-Australian Telescope, the 3.6-m telescope
at ESO and the 4-m telescope at CTIO. The project made use of support
software and data analysis facilities provided by the Starlink Project
which is run by CCLRC on behalf of PPARC. We thank the referee, Dr. 
Martin K{\"u}rster, for suggesting several improvements to the paper. 
ACC acknowledges the support of a PPARC Senior Research Fellowship.




\end{document}